\documentclass[useAMS,usenatbib,usegraphicx]{mn2e}
\usepackage{psfig}   
\usepackage{graphicx}
\usepackage{subfig}
\usepackage{amssymb}

\newcommand{\Msun}{\ensuremath{\textrm{ M}_{\odot}}}
\newcommand{\kms}{\ensuremath{\textrm{ km s}^{-1}}}

\newcommand{\nth}{\ensuremath{^{\rm th}}~}

\title[Testing the initial conditions and dynamical evolution of star clusters using Gaia]{Testing the initial conditions and dynamical evolution of star clusters using Gaia - I}

\author[]{ Richard~J.~Allison$^1$\thanks{E-mail: allison@uni-heidelberg.de}
  \vspace*{0.1cm}\\ $^1$ Institut f\"{u}r Theoretische Astrophysik, Zentrum f\"{u}r Astronomie der Universit\"{a}t Heidelberg, Albert-Ueberle-Str. 2, 69120 Heidelberg, Germany \\}

\begin{document}

\date{}
                             
\pagerange{\pageref{firstpage}--\pageref{lastpage}} \pubyear{2010}

\maketitle

\label{firstpage}

\begin{abstract}

  We investigate how the properties of escaping stars are related to
  the initial conditions of their birth clusters. We find that the
  number of escaping stars, their spatial distribution, and their
  kinematics show a dependence on the initial conditions of the host
  cluster (substructure and virial ratio).  Thus the properties of
  escaping stars can be used to inform us of the initial conditions of
  star formation, and also provide a window into the dynamical history
  of star clusters.  The ESA Gaia mission will make observations of
  the positions and proper motions of these escaping stars at the
  required accuracy to allow us to investigate the dynamical evolution
  of local star clusters and provide an important insight into the
  initial conditions of star formation.

\end{abstract}


\section{Introduction}
\label{sec:intro}

How stars form is one of the outstanding questions in astronomy.  In
particular, it is not clear if stars typically form in clusters
\citep[e.g.,][]{lada03}, in particular in a quasi-static way
\citep{tan06}, or form in a hierarchical distribution
\citep{gutermuth05,bressert10} which can then violently dynamically
evolve into dense star clusters
\citep{allison09,allison10,portegies_zwart10}.  Dynamical interactions
rapidly erase initial substructure and any memory of the initial
conditions, and so deducing the initial conditions from dynamically
evolved clusters and distinguishing these possibilities is difficult.

Constraining the initial conditions of star clusters is important as
their initial state directly influences the early dynamical evolution
of the cluster, such as its ability to remove any initial structure
\citep{goodwin98,goodwin04,smith11}, dynamically mass segregate
\citep{allison09,allison10,moeckel09b,moeckel10}, and process its
binary population \citep{kroupa95,parker09,parker11,moeckel10}.
However, it is difficult to distinguish between a cluster that has
evolved dynamically to be, for example, mass segregated, and one which
was mass segregated at its formation (primordial). Therefore it is
important to be able to estimate whether a cluster is dynamically
evolved or not; but this is difficult to measure as most potential
indicators of dynamical evolution (e.g. mass segregation) can have a
primordial origin. Escaping stars, however, offer a unique measure of
the past dynamical state of a star cluster. Clusters with a violent
past could be distinguishable from clusters with a relatively quiet
history from their escaper population.

The ESA Gaia mission will conduct an all-sky astrometric and
spectrophotometric survey of $\sim 10^9$ point sources between 6\nth
and 20\nth magnitude. The Gaia mission will be able to measure the
distance, position and velocity distribution of the local Galactic
stellar population to high accuracy. Therefore Gaia will not only find
co-moving groups of stars (dispersing clusters), but it will be able to
identify escaping stars from clusters and measure their velocities at
large (angular and physical) distances from their birth cluster.

In this letter we show how escaping stars might be used to investigate
the dynamical history of star clusters and the conditions of their
formation. In Section~2 we describe our initial conditions for
relatively active and relatively quiet clusters.  In Section~3 we show
that the escaper populations from the two clusters are very different
and that these differences should be observable by Gaia.  We discuss
the implications in Section~4, before concluding in Section~5.

\section{Initial Conditions}
\label{sec:init}

We perform $N$-body simulations of young star clusters with two
different sets of initial conditions. The initial conditions are
designed to produce one class of cluster that undergoes rapid and
violent dynamical evolution, and another class that has a relatively
quiescent dynamical history.  However, both produce a
central `cluster' that appear very similar to one another after a
few~Myr.

The initial conditions vary the initial amount of structure and the
initial virial ratio. The initial spatial distribution of stars is
created by using a fractal distribution, as described in detail in
\citet{goodwin04}. Using a fractal distribution allows a simple
procedure with which structured distributions can be created and
reproduced. The method also benefits from having only a single
parameter to control the amount of substructure -- the fractal
dimension ($D$). The fractal dimension paramaterises the structure
such that a lower value implies more substructure (i.e. more clumpy),
and $D=3.0$ is a random distribution. The initial velocities are set
as described in \citet{allison10} and are scaled to a global virial
ratio ($Q$). We define the viral ratio as $Q=T/|\Omega|$ (where $T$
and $\Omega$ are the kinetic and potential energies of stars,
respectively), such that a virialised cluster has $Q=0.5$.

In this study we investigate the effect of initial structure and
virial ratio on the properties of escaping stars. We present two
different initialisations: 1) an initially cool and highly structured
cluster ($D=1.6$, $Q=0.3$); and 2) an initially virialised and less
structured cluster ($D=2.6$, $Q=0.5$), labelled $D$1.6$Q$0.3 and
$D$2.6$Q$0.5, respectively. Each simulation contains 1000 stars, has
an initial maximum radius of 1 pc, includes no primordial binaries or
gas, and a three-part power law is used for the stellar initial mass
function \citep[IMF,][]{kroupa02},

\begin{equation}
  N(M) \propto \left\{
  \begin{array}{r}
    M^{-0.3} \quad m_0 \leq M/\Msun < m_1, \\
    M^{-1.3} \quad m_1 \leq M/\Msun < m_2, \\
    M^{-2.3} \quad m_2 \leq M/\Msun < m_3, \\ 
  \end{array}
  \right.
\end{equation}

\noindent with $m_0=0.08\Msun$, $m_1=0.1\Msun$, $m_2=0.5 \Msun$ and
$m_3=50\Msun$. No stellar evolution is included because of the short
duration of the simulations ($\sim 4$ Myr).  We use the {\sc starlab}
$N$-body integrator {\sc kira} to run our simulations
\citep{portegies_zwart01}. We analyse 50 realisations for each initial 
condition.

For this initial study we have simplified the initial conditions.
Complications such as initial binaries, tidal fields, natal gas, etc
have been ignored to allow a simple numerical experiment to be
performed.  We also ignore various observational selection effects and
assume perfect knowledge of the four dimensions of phase space that
Gaia will best determine, viz. positions on the sky, and proper
motions.  Gaia itself, as well as follow-up observations will also
determine the distance and radial velocities of stars, but to a lower
accuracy.

\section{The variance of properties measurable by Gaia}
\label{sec:results}

Here we investigate two properties of escaping stars that are 
measurable by Gaia:

\begin{enumerate}
\item The number of escaping stars and their spatial distribution around 
their parent cluster;
\item The kinematics of the escaping stars. 
\end{enumerate}

We will show that these properties are dependent on the initial
conditions, and therefore contain much information on the initial
conditions of star formation.

Escaping stars are defined using three criteria:

\begin{enumerate}
\item The magnitude of the velocity is greater than the stars current 
escape speed;
\item The magnitude of the radial component of the velocity vector is 
larger than the tangential component;
\item The star is outside some 'escaper' radius; which is set at two times 
the half mass radius.
\end{enumerate}

\subsection{Spatial distribution of ejected stars}
\label{ssec:spat_dist}

\begin{figure*}
  \begin{center}
\subfloat[]{\label{sfig:gp.faa11a001}
  \includegraphics[scale=0.5,angle=270]{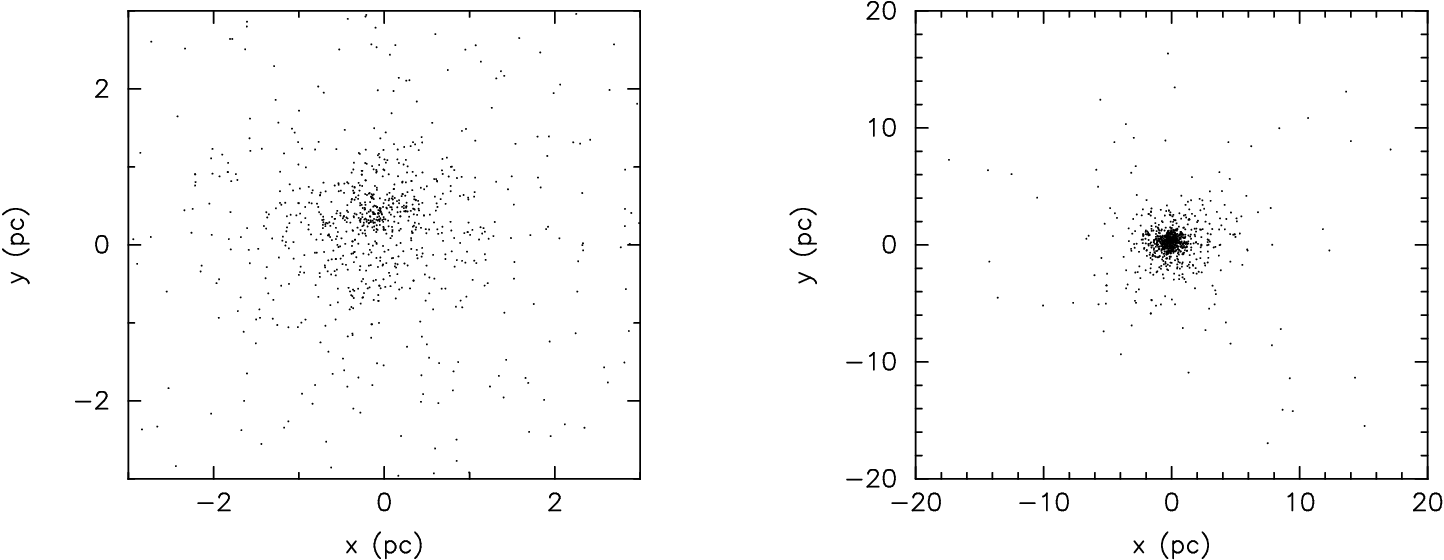}}
\\
\subfloat[]{\label{sfig:gp.fac31a001}
  \includegraphics[scale=0.5,angle=270]{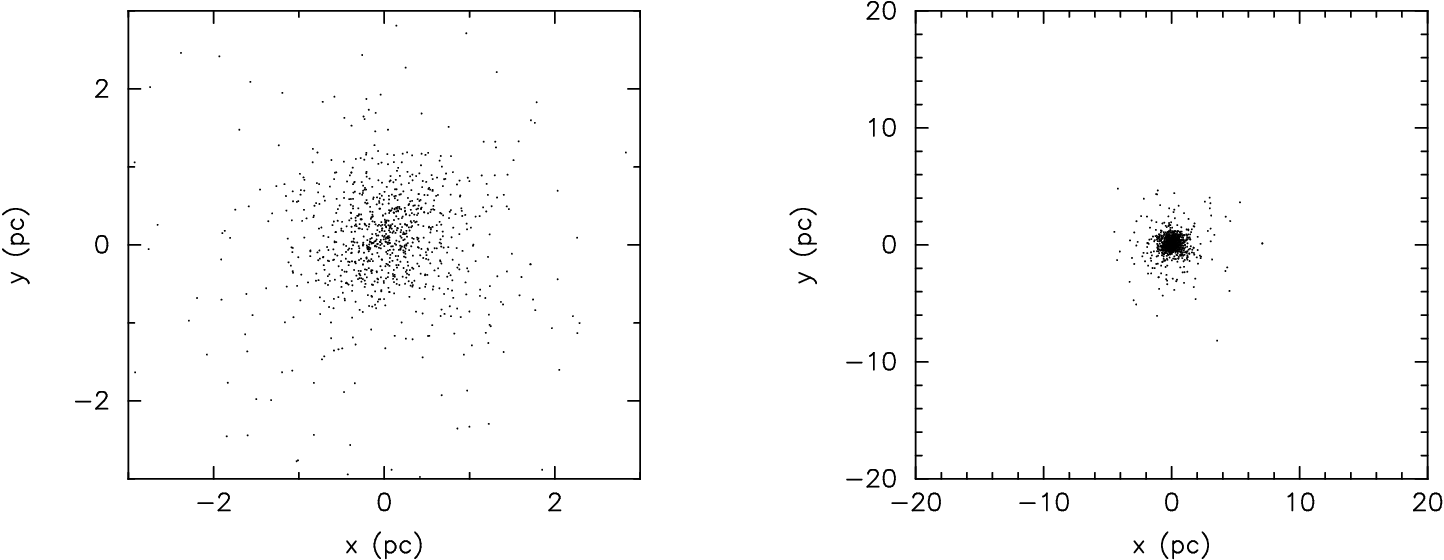}}
  \end{center}
  \caption[]{
The two-dimensional projection of (a) an initially cool and
clumpy cluster ($Q=0.3$ \& $D=1.6$), and (b) an initially virialised
and smoother distribution ($Q=0.5$ \& $D=2.6$) at 4 Myrs. The plots on
the left show the central regions on 2 pc scale, the differences
between the spatial distributions are not obvious. However, the right
side plots show a zoomed out view (20 pc), here the differences are
clear. The initially clumpier cluster has a much more pronounced
population of escaping stars, caused by its more energetic dynamical
evolution.
  \label{fig:spat_dist}}
\end{figure*}

Our two clusters evolve in very different ways for the first $2$~Myr.
The clumpy and cool cluster ($D$1.6$Q$0.3) undergoes a dense collapse
phase and re-expansion before reaching a Plummer-like equilibrium
state by $4$~Myr as described in \citep{allison10}.  The smoother,
virialised cluster ($D$2.6$Q$0.5) gently relaxes to a Plummer-like
equilibrium in roughly a crossing time \citep[see also,][]{allison10}.
The final states of the clusters are illustrated in
Figure~\ref{fig:spat_dist}.  The left-hand panels show the inner 3~pc
radius region of typical examples of the two clusters after 4~Myr, as
can be seen the spatial distribution of the two clusters is very
similar and memory of their initial conditions has been erased.
However, the right-hand panels show the spatial distribution on a
20~pc radius scale in which there is clearly a difference: the
$D$1.6$Q$0.3 cluster having more, and more widely distributed, escaping
stars.

Clumpy and cool clusters undergo a much more dynamical evolution and
smooth and relaxed clusters. Previous work on the effect of initial
spatial distributions has shown that clusters which are initially
clumpier tend to have a more violent, dynamical evolution, undergoing
mass segregation \citep{allison09,allison10,moeckel09}, forming
trapezium-like systems \citep[high-order multiples containing
high-mass stars;][]{allison11}, and even influencing the binary
properties \citep{parker11}.  In smoother, relaxed clusters these
effects are much less pronounced, and it is likely that features such
as trapezium systems and mass segregation cannot have a dynamical
origin \citep{bonnell98,allison10} and so must have a `primordial'
origin in the nature of star formation itself.

The spatial distribution of stars in the central regions of these
initially very different clusters (left-hand plots in
Figure~\ref{fig:spat_dist}) show few differences.  But if we move out
beyond the pc-scale into the 10s pc-scale (right-hand plots) we can
immediately observe that these clusters have had a very different
dynamical history.  Essentially, the differences in the numbers,
distance and velocities of escapers is a window into the dynamical
history of the cluster.  More and faster escapers means that a cluster
is dynamically older and has had more two-body encounters within it to
eject stars.

The $D$1.6$Q$0.3 cluster shown in Figure~\ref{fig:spat_dist} (top
panels) has 80 escaping stars as projected on the sky (in three
dimensions it has actually 149 escaping stars).  This is compared to 15
escapers in projection (30 in three dimensions) for the $D$2.6$Q$0.5
cluster (bottom panels).  These results are typical throughout our
ensembles, with clumpy, cool clusters producing an average of 57
escapers, and smooth, relaxed clusters producing only 12 in
projection (115 and 20 in three dimensions, respectively).

\subsection{Kinematics}
\label{ssec:kine}

\begin{figure}
  \begin{center} 
\includegraphics[scale=0.3,angle=270]{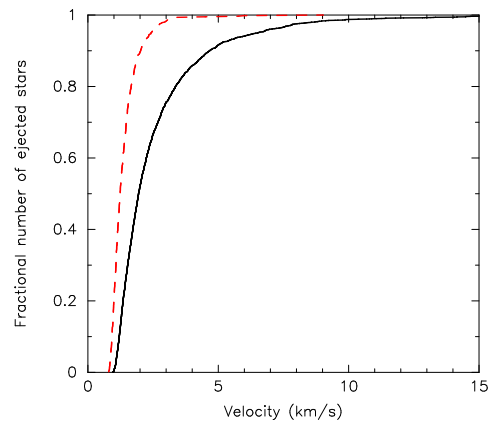}
  \end{center}
  \caption[]{
The cumulative distribution function of the velocities of escaping
stars for the cool and clumpy (black, solid line) and virialised and
smoother (red, dot-dashed line) clusters at 4 Myrs. The differences between the
velocity distributions are caused by the dynamical evolution of the
clusters. The cool and clumpy cluster is able to produce faster moving
escaping stars through its dynamic early evolution and eject stars at
high velocities through binary star formation. The virialised smoother
cluster has a more quiescent early evolution, and does not produce a
population of fast moving escaping stars.
  \label{fig:vcdf.faa1_c3}}
\end{figure}

The differences between the spatial distributions of the escaping
stars from the two clusters is also seen in the kinematics of
their escaping stars. Figure \ref{fig:vcdf.faa1_c3} shows the
cumulative distribution of the escaping stars from the $D$1.6$Q$0.3
cluster (black solid line) and the $D$2.6$Q$0.5 (red dashed line). There
is a clear difference between the kinematics of the two sets of
escaping stars. The $D$2.6$Q$0.5 cluster shows a small spread of escaper
velocities, with 90 per cent of the ejected stars within a 0.5\kms
spread around 1.5\kms, with a median escaper velocity of 1.2\kms. The
maximum escaper velocity from this cluster is $<10\kms$. While the $D$1.6$Q$0.3
cluster, has a larger 2\kms spread for 90 per cent of escapers (with a
median velocity of 2\kms), and has a much more pronounced high
velocity tail. 10 per cent of escaping stars have velocities over
5\kms and the maximum escaper velocity is 15\kms.

It should be noted that there is some degeneracy between the distance
of an escaping star from the cluster and its ejection velocity which
Gaia can break.  An ejected star at 10~pc from the cluster may have
been ejected 4~Myr ago at 2.5~km~s$^{-1}$, or 1~Myr ago at
10~km~s$^{-1}$.  The difference between these two situations is
important as a 10~km~s$^{-1}$ requires a much stronger interaction
(probably with a massive binary) than a 2.5~km~s$^{-1}$ ejection event
(see below).

The velocity cumulative distribution functions shown in
Figure~\ref{fig:vcdf.faa1_c3} can be separated into two populations,
high- ($\gtrsim 5\kms$) and low-velocity ($\lesssim 5\kms$). The high
velocity tail is populated by stars ejected by high-mass binary systems.
This population is mostly missing from the $D$2.6$Q$0.5 clusters because
they are not able to easily form these binary systems (the clusters
are initially single stars). While the more dynamic early evolution of
the $D$1.6$Q$0.3 clusters allows the formation of high-mass binary systems
\citep[e.g.,][]{allison11}. In clusters with a primordial binary
population the formation of high-mass binary systems is easier, as
forming binaries through dynamics is a difficult process
\citep{kroupa95}. 

If primordial massive binaries are present then a high-velocity tail
should also be present even in quiescent clusters, although we would
still expect a smaller number of ejections as the number of encounters
with the massive binary should be lower (due to the younger dynamical
age of the quiescent cluster).  We will return to this in a later
paper when we examine the effect of primordial binaries.

\section{Discussion}
\label{sec:discussion}

We have investigated the influence of the initial conditions of star
cluster on the properties of escaping stars. We have shown that after
$4$~Myr of evolution clusters with very different initial conditions
in both substructure and virial ratio) appear to have similar spatial
distributions on pc-scales. However, on larger scales ($\sim
10$pc) the spatial distributions of the clusters are very different
(Figure~\ref{fig:spat_dist}). Clusters which are initially cool and
clumpy have significantly more escaping stars, and these escapers have
a higher average velocity, and a pronounced high-velocity tail (Figure
\ref{fig:vcdf.faa1_c3}).  Therefore, one can learn about the dynamical
evolution and initial conditions of star clusters, by observing the
population of escaping stars around star clusters.

Disentangling the effects of dynamical evolution and birth properties
in star clusters is a complex problem. For example, can one use young
star clusters ($< 1$ Myrs) to investigate the primordial binary
population? Dynamical simulations have shown that star clusters
process their primordial binary population as they dynamically evolve,
therefore it is important that we are able to estimate how {\it
  dynamically evolved} a cluster is if we are to infer information
about its primordial state. However, can we distinguish between a
primordial binary population that {\it looks} dynamically processed
and a population that actually has been? Likewise with mass
segregation - how does one distinguish between `primordial' and
dynamical mass segregation? Clearly, dynamical mass segregation {\it
  must} only occur in dynamically evolved clusters (as this occurs on
the relaxation timescale). Therefore, mass segregated clusters that
are not dynamically evolved but are mass segregated are primordially
mass segregated.

Measuring the true dynamical age of a star cluster is difficult. Clusters
such as the Orion Nebula Cluster have {\it current} relaxation times
of $\sim 50$ Myrs, which implies that they are dynamically very young,
having an age $\approx 1/10$ its relaxation time. However, dynamical
simulations have shown that current cluster timescales can be
misleading. Clusters that undergo an initial collapse and re-expansion
have had much shorter relaxation times in their past, and can therefore
be dynamically more evolved than they appear when using their current
relaxation times \citep{bastian08,portegies_zwart08,allison10}.

Escaping stars are therefore a very useful population to investigate,
as they are dynamical in origin - there is no obvious primordial
process that would produce a significant population of escaping stars
from a cluster. Thus, measuring the population of escaping stars from
a star cluster (and their properties, e.g. velocities) tells us about
the dynamical history of the cluster. This work is the first step into
looking at how the escaping population is related to initial
conditions of star clusters.

Stars that escape from clusters retain information about the dynamical
state of the cluster at the time that they are liberated. Thus, the
kinematics of escaping stars can provide us with information about the
state of their host cluster at the time of their escape. For example,
high-velocity escapers are an indication of either the formation of a
binary or an interaction with one. The kinematics of a star ejected by
a binary can inform us of the properties of the binary, and the
velocity and distance of the escaping star allow us to constrain a
time at which the interaction occurred \citep[see
e.g.,][]{tan04,zapata09}. In particular, examining the high-velocity
escapers produced by interactions with massive binaries and
Trapezium-like systems may constrain the formation time of those
systems and distinguish between primordial and dynamical models for
their formation.

The Gaia mission provides the first opportunity to examine the
dispersed population around star clusters.  As it is all-sky it will
find stars at great (angular) distances from clusters and the
additional kinematic and distance information provided will allow
stars to be associated with their birth cluster even if they have
travelled a large distance on the sky.

What we have not addressed in this letter are the observational
limitations of the Gaia data.  Sky position and proper motions will be
obtained to an accuracy of positions and proper motions are expected
to be $\sim 200\mu$arcseconds and $\sim 150\mu$arcseconds,
respectively, for 20th magnitude objects \citep{lindegren10}.  This
means that Gaia will have accurate positions and proper motions for
G-dwarfs at 1~kpc, but for M-dwarfs only out to a few 100~pc.
Therefore the masses of escapers from more distant clusters will be an
important observational limitation.  We will return to this in later
work.

\section{Conclusions}
\label{sec:conc}

We have investigated how the properties of escaping stars are related
to the initial conditions of their host clusters. We find that the
number of escaping stars and their spatial distribution/kinematics
show a dependence on the initial conditions of the host cluster
(substructure and virial ratio). Escaping stars from initially cooler
and clumpier clusters tend to be more numerous, and have a larger
spread in velocities. While virialised and smoother clusters are less
efficient at producing escaping stars. These simulations have been
carried out with simple initial conditions (i.e. no binaries, gas,
tidal fields), although there is no reason to think that the addition
of more complex initial conditions/processes will significantly alter
the results as the results depend on the dynamical age of the cluster.
But following work will investigate the influence of more complicated
initial conditions. The properties of escaping stars can be used to
inform us of the initial conditions of star and cluster formation, and
can provide us with a window into the dynamical history of star
clusters.

The ESA Gaia mission will provide the all-sky coverage and dynamical
information required to investigate escaping populations from star
clusters and provide an important insight into the initial conditions
of star formation.

\section{Acknowledgments}

RJA would like to thank Simon Goodwin, Simon Portegies Zwart, Richard
Parker and Andreas Koch for useful discussions that improved this
manuscript.  RJA also acknowledges support from the Alexander von
Humboldt Foundation in the form of a research fellowship. This work
has made use of the {\sc Iceberg} computing facility, part of the
White Rose Grid computing facilities at the University of Sheffield.
The simulations reported in this paper also made use of the Kolob
cluster at the University of Heidelberg, which is funded in part by
the DFG via Emmy-Noether grant BA 3706.


\bibliography{gaia_bib}
\bibliographystyle{mn2e}

\end{document}